\documentclass[12pt,preprint]{aastex}

\lefthead{Goss \& Brogan}
\righthead{Eye of the Tornado}

\begin{document}
\newcommand\HII{H\,{\sc ii}}
\newcommand\HI{H\,{\sc i}}
\newcommand\OI{[O\,{\sc i}] 63 $\mu$m}
\newcommand\CII{[C\,{\sc ii}] 158 $\mu$m}
\newcommand\CI{[C\,{\sc i}] 370 $\mu$m}        
\newcommand\SiII{[Si\,{\sc ii}] 35 $\mu$m}
\newcommand\Hi{H110$\alpha$}
\newcommand\He{He110$\alpha$}
\newcommand\Ca{C110$\alpha$}
\newcommand\kms{km~s$^{-1}$}
\newcommand\cmt{cm$^{-2}$}
\newcommand\cc{cm$^{-3}$}
\newcommand\Blos{$B_{los}$}
\newcommand\Bth{$B_{\theta}$}
\newcommand\Bm{$B_{m}$}
\newcommand\Bvm{$\mid\vec{B}\mid$}
\newcommand\BS{$B_S$}
\newcommand\Bscrit{$B_{S,crit}$}
\newcommand\Bw{$B_W$}
\newcommand\mum{$\mu$m}
\newcommand\muG{$\mu$G}
\newcommand\mjb{mJy~beam$^{-1}$}
\newcommand\jb{Jy~beam$^{-1}$}
\newcommand\dv{$\Delta v_{FWHM}$}
\newcommand\va{$v_A$}
\newcommand\Np{$N_p$}
\newcommand\np{$n_p$}
\newcommand\pp{$^{\prime\prime}$}
\newcommand\km{km~s$^{-1}$}
\newcommand\h{^{\rm h}}
\newcommand\m{^{\rm m}}
\newcommand\s{^{\rm s}}

\title{VLA Observations of the ``Eye of the Tornado''- the High Velocity \HII\/ 
Region G357.63$-$0.06} 
 
\author{ C. L. Brogan \& W. M. Goss}
\affil{National Radio Astronomy Observatory, Very Large Array, P. O. Box 0,
Socorro, New Mexico 87801, USA}
 
\email{cbrogan@aoc.nrao.edu}
 
\begin{abstract}

The unusual supernova remnant candidate G357.7$-$0.1 and the compact source G357.63$-$0.06
have been observed with the Very Large Array at 1.4 and 8.3 GHz.  The H$92\alpha$ line (8.3
GHz) was detected from the compact source with a surprising velocity of $\sim -210$ \km\/
indicating that this source is an \HII\/ region, is most likely located at the Galactic
center, and is unrelated to the SNR.  The \HI\/ absorption line (1.4 GHz) data toward these
sources supports this picture and suggests that G357.7$-$0.1 lies farther away than the
Galactic center.

\end{abstract}


\section{INTRODUCTION}

The unusual and intense radio source G357.7$-$0.1, the Tornado, has been the subject of
controversy since it was first recognized by \citet{Milne1970} \citep[also
see][]{Caswell1980,Caswell1989, Shaver1985a,Weiler1980}.  The Tornado has a non-thermal spectrum
($-0.5\la\alpha\la -1.0$) and is polarized \citep{Stewart1994}.  The morphology of the Tornado is
unique among galactic sources and it has been suggested that the source is extragalactic, a SNR
(supernova remnant), or a member of a new class of galactic head tail sources related to W50 and
SS443 \citep{Becker1985,Helfand1985}.  The discovery of a collisionally pumped OH (1720 MHz) maser
near G357.7$-$0.1 by \citet{Frail1996} has renewed suggestions that the source could, in fact, be a SNR.
Using the Zeeman effect, \citet{Brogan2000} observed a magnetic field of 0.7 $\mu$G in the OH
maser.

The nature of the flat spectrum, compact source G357.63$-$0.06 at the western extreme of the
Tornado has been also controversial.  It has been suggested that the source could contain a
pulsar and might be a pulsar wind nebula physically related to the possible SNR G357.7$-$0.1
\citep{Shull1989}.  These authors raise the possibility that this putative pulsar might be
interacting with the supernova shell and be responsible for the unusual morphology of
G357.7$-$0.1, similar to CTB 80.  Alternatively, \citet{Shaver1985b} have summarized evidence
that the source is an \HII\/ region that may be unrelated to G357.7$-$0.1.  As suggested by
\citet{Brogan2000}, higher resolution radio continuum and recombination line observations of
G357.63$-$0.06 were necessary in order to determine if the source is an \HII\/ region, as had
been suggested by its flat radio spectrum.  In the following, all references to the ``Tail'',
``Head'', and ``Tornado'' refer to the supernova remnant candidate G357.7$-$0.1, while the
``Eye'' refers to the compact source G357.63$-$0.06.  Figure 1 shows a 21 cm image created from
archival VLA data with $\sim 12\arcsec$ resolution in which the Tail and Head of the Tornado and
the Eye are identified for reference.

A search for radio recombination lines (RRL) toward G357.63$-$0.06 (the Eye) was begun in 2000
using the NRAO\footnote{The National Radio Astronomy Observatory (NRAO) is a facility of the
National Science Foundation operated under a cooperative agreement by Associated Universities,
Inc.}  Very Large Array (VLA).  After several unsuccessful searches covering the velocity range of
the OH maser ($-$12.3 \km\/) in 2000, we learned of a detection of a Br$\gamma$ line at high
negative velocities using the Anglo-Australian Observatory by \citet{Burton2002}.  We subsequently
detected the H$92\alpha$ line at the surprising velocity of $-210$ \km\/ with the VLA in 2001.  In
order to better constrain the distances of the Tornado and the Eye, in 2002 we also observed the
\HI\/ line in absorption with the VLA toward these sources.

\section{OBSERVATIONS}

The VLA observations of 2000 and 2001 of the H$92\alpha$ (8.309 GHz) line are summarized in
Table 1.  The initial observations in June and July 2000 only cover the velocity range $-100$
to 60 \km\/, and are centered at $-12.3$ \km\/, the velocity of the OH maser detected by Frail
et al.\ (1996).  The wider band width observations of late 2000 (also centered at $-12.3$
\km\/) cover the velocity range of $-195$ to +190 \km\/.  The data obtained in 2001 were made
with a more favorable center velocity ($-200$ \kms\/) and include the range $-400$ to 0 \km\/.
We also observed the \HI\/ line at 1.42 GHz in absorption toward the Tornado, and the
parameters of these observations are also given in Table 1.  In addition, the parameters of the
archival 1.4 and 4.9 GHz continuum data used in our analysis are also shown in Table 1.  All of
these data were reduced in the usual fashion using the AIPS (Astronomical Image Processing
Software) package.  The H$92\alpha$ and \HI\/ spectral line data were Hanning smoothed and
their spectral resolutions can also be found in Table 1.  Subsequent spectral line Gaussian
fitting and spatial integration were carried out using GIPSY\footnote{The Groningen Image
Processing System (GIPSY) is documented at {\tt http://www.astro.rug.nl/${\mathtt
\sim}$gipsy/}.}.

\section{RESULTS}

In Figure 2, we show VLA 8.3 GHz continuum images of the Head (G357.7$-$0.1) and Eye
(G357.63$-$0.06) from the combined 2000 data and line free channels of the 2001 data.  The Head
image was created from UV spacings less than 100 k$\lambda$ (to achieve good image quality and
sensitivity to the extended emission) and its resolution is $14\farcs 3\times 6\farcs 8$.  The
image of the Eye has a resolution of $3\farcs 1\times 2\farcs 1$ and it was obtained by
including all of the longer baselines up to 350 k$\lambda$.  The position of the Eye is RA
$=17{\rm^h}40{\rm^m}05\fs68\pm 0\fs01$, DEC $=-30\arcdeg58\arcmin56\farcs1\pm 0\farcs1$
(J2000).  It is clear from this high resolution image that G357.63$-$0.06 is not spherically
symmetric or of uniform brightness, and its morphology is most likely shell type (see also
Burton et al.  2002).  Both images yield a total flux and size for the compact source,
G357.63$-$0.06 of $\sim 88\pm 2$ \mjb\/ and $\sim 6\arcsec$, respectively.
The results from the 8.3 GHz images along with the parameters derived from the line-free 1.4
GHz \HI\/ data, archival 1.4 GHz data (also see Fig.  1), and archival 4.9 GHz data are listed
in Table 2.

For the 8.3 GHz H$92\alpha$ line data of 2001 alone (beam $14.9\arcsec\times 6.6\arcsec$),
the rms noise is 0.4 \mjb\/ and the peak of the G357.63$-$0.06 continuum is 57.6 \mjb\/.
In Figure 3, we show the H$92\alpha$ spectrum toward the peak of G357.63$-$0.06.  Based on
a Gaussian analysis, the line parameters are:  $v= -210\pm 3$ \km\/, $\Delta V= 36\pm 7$
\km\/, and the line to continuum ratio is $0.042\pm 0.007$.

Figure 4 shows integrated, continuum weighted \HI\/ spectra from the Head and Eye of
the Tornado.  Both spectra clearly show the 3 kpc arm of the Galaxy at $\sim -60$ \kms\/, as
well as a number of other distinct features at about $-94$, $-30$, $-10$, $+10$, and $+45$ \kms\/.
Interestingly, the \HI\/ spectrum toward the Head also shows weak absorption near $-210$
\kms\/ (inset in Fig.  4).  The \HI\/ spectrum toward the Eye itself is of insufficient
signal to noise to detect the weak $-210$ \kms\/ feature.

\section{DISCUSSION}

\subsection{Distances}

It is certain that both the Eye and the body of the Tornado are at a distance greater than 5 kpc
since \HI\/ spectra toward both sources (Fig.  4) clearly show the 3 kpc arm of the Galaxy at
$\sim -60$ \km\/ \citep[also see][]{Rad1972}.  The detection of a RRL toward the Eye is
unequivocal evidence that G357.63$-$0.06 is an \HII\/ region.  The surprisingly high velocity of
this \HII\/ region $\sim -210$ \kms\/ implies that it is kinematically well separated from the OH
(1720 MHz) maser discovered by \citet{Frail1996} located $\sim 1.4\arcmin$ to the NE of the Eye at
a velocity of $v= -12.3$ \km\/ (see Fig.  2).  If the OH (1720 MHz) maser is associated with the
Head of the Tornado as inferred by Frail et al., the location of G357.63$-$0.06 along the symmetry
axis of the Tornado (see Fig.  1) must be viewed as a chance superposition along the line of sight
in agreement with the suggestion of \citet{Stewart1994}.

It is likely that G357.63$-$0.06 lies close to the Galactic center, most likely in the Nuclear
disk as this is the only place where such high negative Galactic rotation velocities ($\sim -200$
\kms\/) are found \citep[e.g.][]{Burton1978,Liszt1980,Dame1987, Dame2001}.  If we assume this is
the case, the minimum projected distance of this \HII\/ region from Sgr A is $\sim 350$ pc.
Similar velocities of $\sim -200$ \kms\/ have been observed $\sim 1\arcdeg$ NE of G357.63$-$0.06
towards the SgrE \HII\/ region complex via radio recombination lines by \citet{Cram1996} and
\citet{Lockman1996}.  The CO $b-v$ and $\ell-v$ diagrams of the Galactic center region from
\citet{Bitran1997} show a resolved clump of CO emission about $0.5\arcdeg$ in extent at $\ell\sim
357.7\arcdeg$, $b\sim -0.1\arcdeg$ and $v\sim -210$ \kms\/, confirming the presence of molecular
gas in the vicinity of G357.63$-$0.06 \citep[see also][]{Burton2002}.  CO observations by
\citet{Shaver1985b} toward the Eye also show a feature at 9 \km\/, but this component is most
likely related to a foreground molecular cloud, and is not associated with G357.63$-$0.06.

Our new high dynamic range, integrated \HI\/ absorption spectrum towards the Head of the
Tornado (Fig.  4) also reveals for the first time an absorption feature at $\sim -210$ \kms\/.
The detection of this component unambiguously places the Tornado at least as far away as the
Galactic Center distance of 8.5 kpc.  If the OH (1720 MHz) maser velocity ($-12.3$ \km\/) is
assumed for the body of the Tornado, the kinematic (far) distance to both the OH (1720) MHz 
maser and SNR candidate is $\sim 12$ kpc
\citep{Frail1996}.

\subsection{Properties of G357.63$-$0.06}

The measured properties of the Eye at 1.42, 4.86, and 8.31 GHz from this study are shown in
Table 2.  As noted in previous studies the Eye has a relatively flat, but slowly rising
spectrum below 4.9 GHz \citep{Shaver1985b,Burton2002} indicative of an optically
thin \HII\/ region.  Our radio flux measurement at 8.3 GHz is the first to fill the gap between
4.9 GHz and 1.3 mm (upper limit non-detection at 1.3 mm of 1 Jy by \citet{Shaver1985b}).  The
measured 8.3 GHz flux density of 88 mJy is consistent (within the noise) with an optically thin
\HII\/ region spectrum assuming $S_{\nu}\propto\nu^{0.1}$.  Studies of the spectral energy
distribution of G357.63$-$0.06 by \citet{Shaver1985b} and \citet{Burton2002} using
infrared colors (IRAS and MSX) show that there can be very little contribution from dust at 8.3
GHz, as expected, in agreement with our findings.

The derived parameters of the \HII\/ region G357.63$-$0.06 assuming $d=8.5$ kpc, $S=88$ mJy
(8.3 GHz flux density), size=$6\arcsec$, and $Y^+=0.1$ are:  shell radius $Rs=0.2$ pc, electron
density $ne=1.6\times 10^3$ \cc\/, emission measure $EM =1\times 10^6$ pc cm$^{-6}$, excitation
parameter $u=26.9$ pc cm$^{-2}$, mass of ionized gas $M_{\rm HII}\sim 1.2$ M$_{\sun}$, and number
of Lyman continuum photons $Nc=5.7\times 10^{47}$ photons.  The LTE electron temperature of the
ionized gas is $T_e\sim 11,000$ K assuming $Y^+=0.1$.  If we assume that a single star powers
the \HII\/ region given its shell like morphology (Fig.  2), these parameters indicate that the
central star is about O9 or B0 type.  These estimates for the parameters of the Eye are similar
to those listed by \citet{Shaver1985b} and \citet{Burton2002}.

\section{CONCLUSIONS}

We have unambiguously identified G357.63$-$0.06 as a Galactic \HII\/ region via the discovery
of a 8.3 GHz H$92\alpha$ radio recombination line at a velocity of $\sim -210$ \kms\/.  Based
on this unusually high negative velocity along with new \HI\/ absorption data, we show that
this source most likely lies near the Galactic Center with a minimum projected distance of
350 pc.  Using the radio continuum parameters of G357.63$-$0.06, we find that the mass of
the ionized gas is $M_{\rm HII}\sim 1.2$ M$_{\sun}$ and that the number of Lyman continuum
photons is consistent with this source being powered by a O9 or B0 type star.

Assuming that the OH (1720 MHz) maser discovered by \citet{Frail1996} is associated with the
the Tornado (G357.7$-$0.1) SNR candidate, we also find that the Tornado and Eye are well
separated kinematically.  Thus the location of the Eye along the symmetry axis of the Tornado,
while suggestive of a unique interaction given the Tornado's unusual morphology, is simply due
to chance superposition along the line of sight.  Additionally, high dynamic range \HI\/
absorption data towards the body of the Tornado suggests that it too must lie at least as far
away as the Galactic center and most likely at 12 kpc if the OH (1720 MHz) maser velocity is
assumed.

\acknowledgments

We would like to thank J.  Lazendic for sharing with us the Br$\gamma$ velocity of
G357.63$-$0.06 ahead of publication which allowed us to shift our H$92\alpha$ radio
recombination line observations to a more favorable center velocity.

\newpage
 
\section{Figure captions}

\begin{figure}[h!]
\plotone{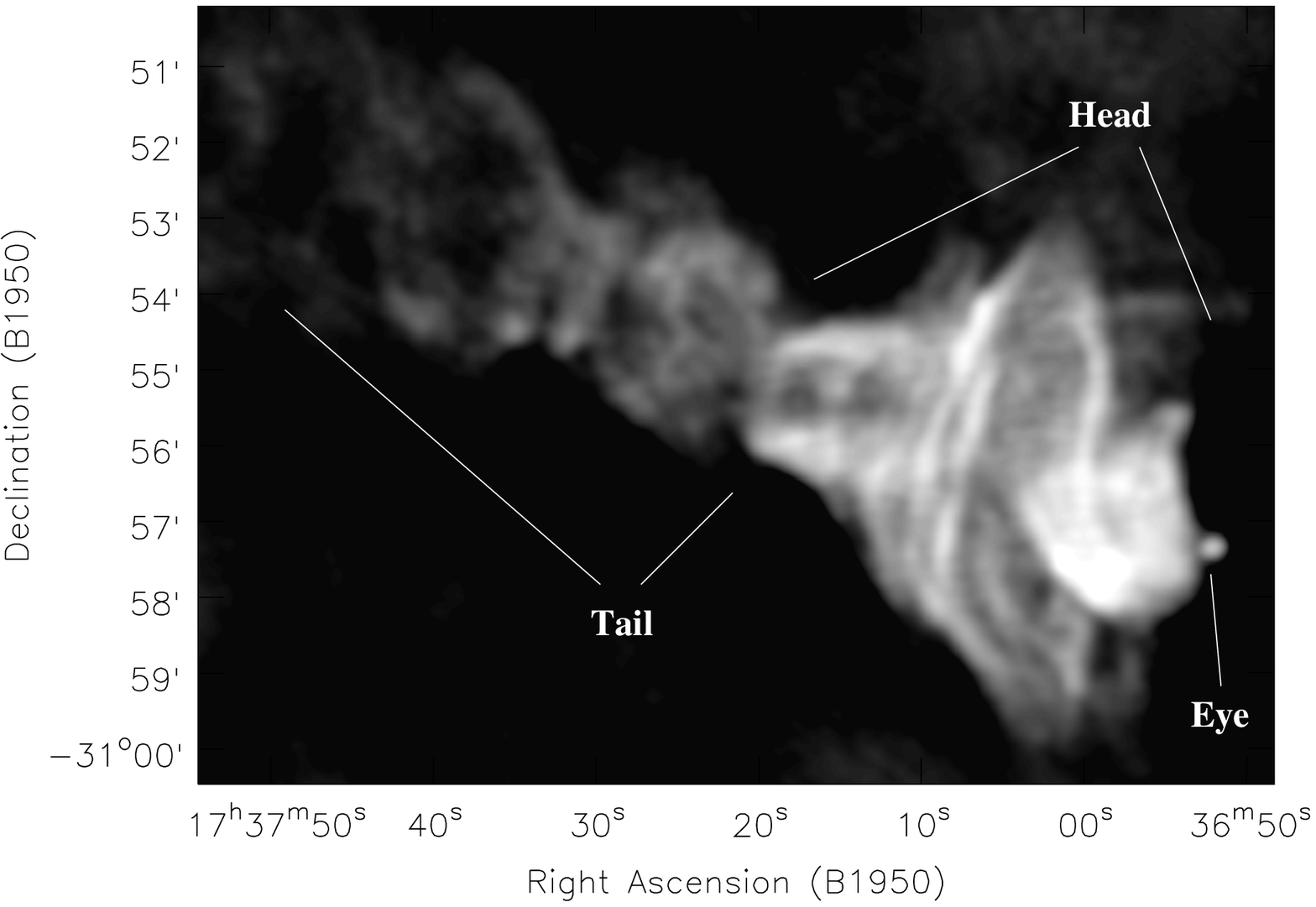}
\caption[]{VLA 1.4 GHz continuum image of the Tornado with $13\farcs 6\times 11\farcs 4$
(P.A.$=-76\arcdeg$) resolution constructed from CnB and DnC configuration archival data.  The
Tail, Head, and Eye of the Tornado are indicated for reference.  The peak flux density in this
image is 230 \mjb\/ and the rms noise is 0.6 \mjb\/.  No primary beam correction was applied 
to this image.}
\end{figure}

\begin{figure}[h!]
\plotone{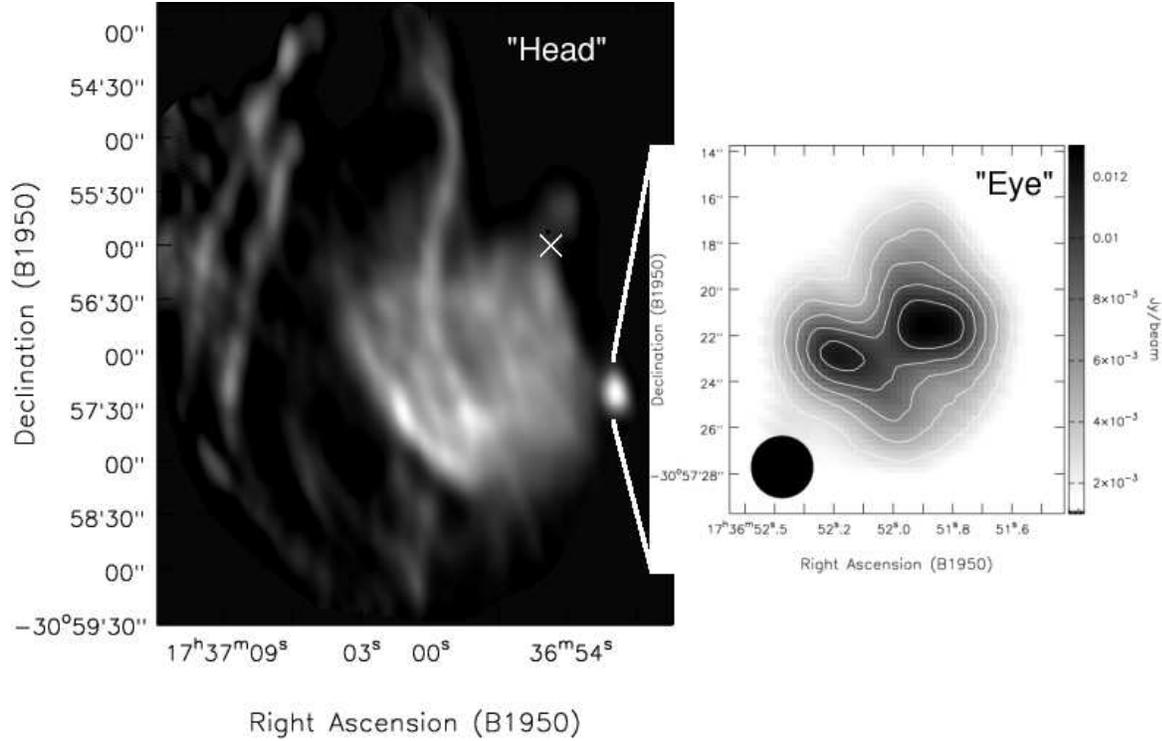}
\caption[]{VLA 8.3 GHz continuum images of the Head and Eye of the Tornado from the combined
2000 and 2001 data.   The resolution of the Head image is $14\farcs 3\times 6\farcs 8$
(P.A.$=11\fdg 7$), while the resolution of the Eye image is $3\farcs 1\times 2\farcs 1$
(P.A.$=14\fdg 6$).  The Head image has been corrected for the shape of the primary beam and has
a peak flux density of 61 \mjb\/ and rms noise of 0.3 \mjb\/.  The peak flux density of the Eye
image is 13 \mjb\/ and the rms noise is 0.2 \mjb\/.  The contours on the Eye image are at 3, 5,
7, 9, and 11 \mjb\/.  The $\times$ symbol marks the location of the $-12.3$ \kms\/ OH (1720
MHz) maser \citep{Frail1996}.}

\end{figure}

\begin{figure}[h!]
\plotone{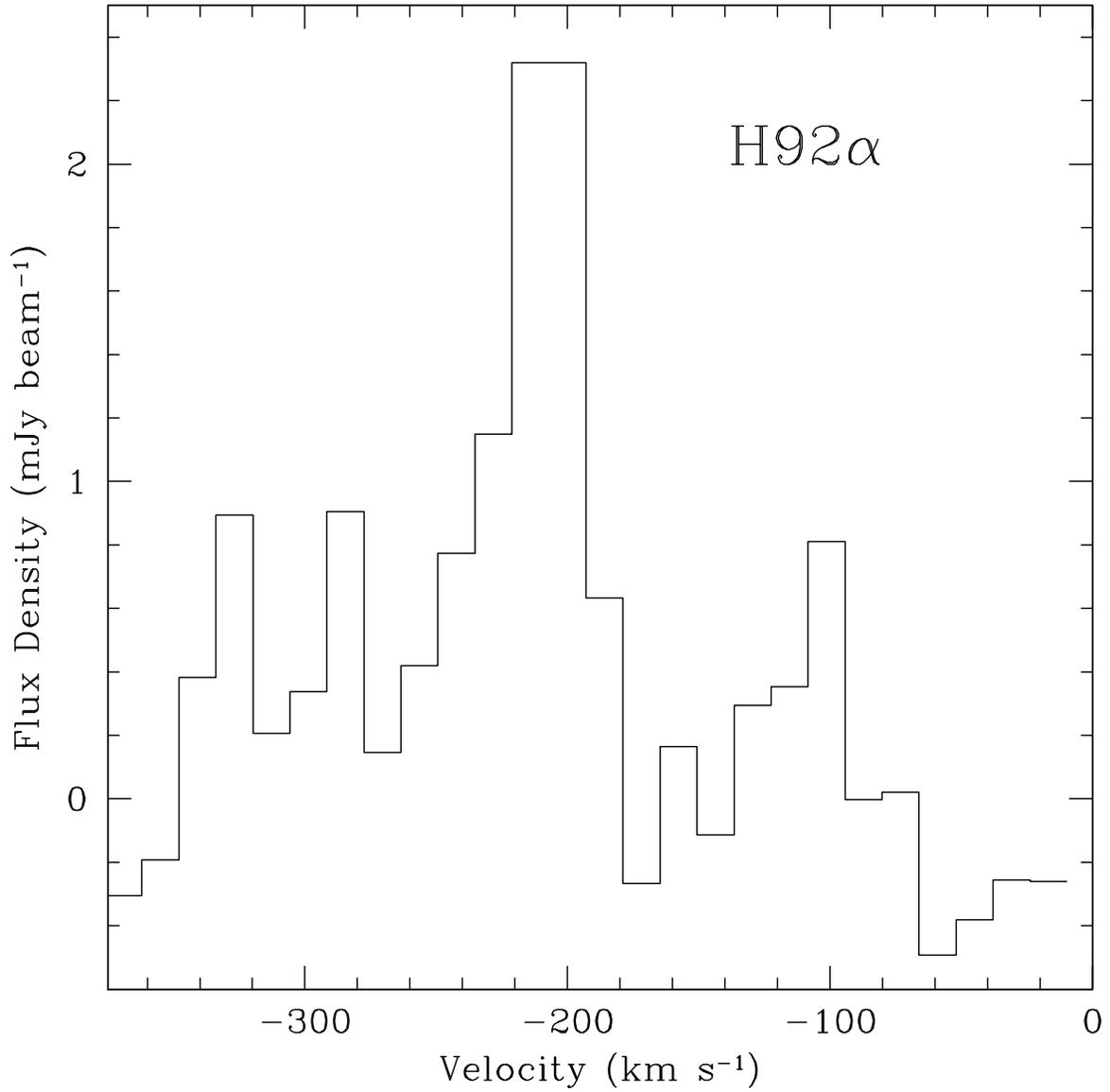}
\caption[]{H$92\alpha$ radio recombination line profile toward the Eye.  These data have
$14\farcs 9\times 6\farcs 6$ spatial resolution and the channel separation is 14 \kms\/.  The
rms noise in the spectrum is 0.3 \mjb\/.  The 8.3 GHz continuum peak at this
position is 57.6 \mjb\/ and the continuum rms noise is 0.2 \mjb\/.}
\end{figure}

\begin{figure}[h!]
\plottwo{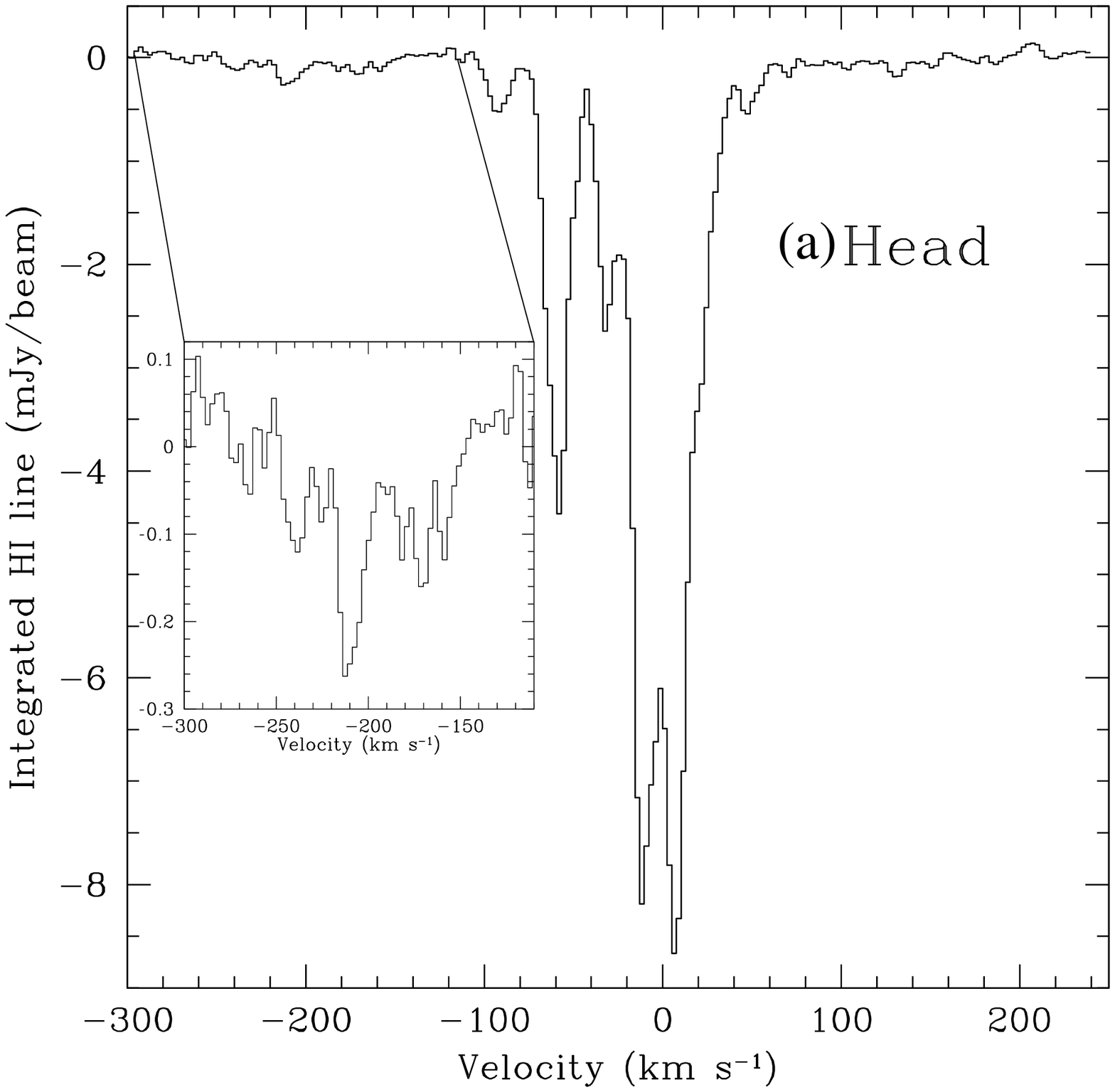}{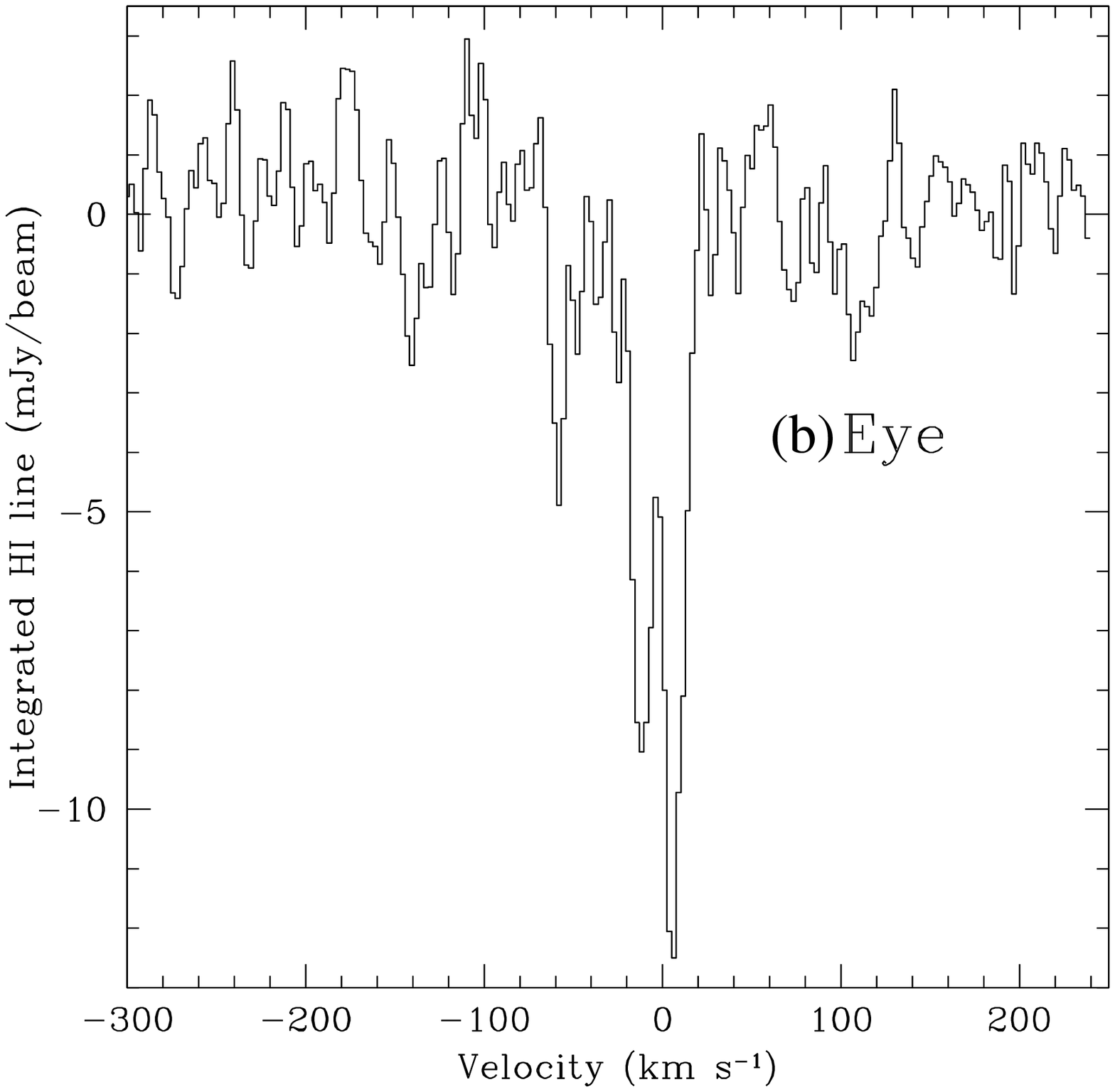}
\caption[]{Integrated, continuum weighted \HI\/ profiles toward the 
(a) Head and (b) Eye of the Tornado. The inset on the Head profile (a) shows an expanded view 
of the weak $-210$ \kms\/ \HI\/ component. After integration, the rms noise in the 
Head spectrum is $\sim 0.05$ \mjb\/. For the Eye spectrum (b) the rms noise is $\sim 1.2$ \mjb\/.} 
\end{figure}

\newpage

\begin{deluxetable}{lccccc}
\small
\tablewidth{36pc}
\tablecaption{VLA Tornado Observational Parameters\label{tab1}}
\tablecolumns{6}
\tablehead{
\colhead{Date}  & \colhead{Config.} & \colhead{Center Velocity} & \colhead{Bandwidth} & \colhead{Channel Width} & \colhead{Time~$^a$} \\
  &   & \colhead{(km s$^{-1}$)} & \colhead{(km s$^{-1}$)} & \colhead{(km s$^{-1}$)} & \colhead{(hours)}}     
\startdata
\cutinhead{8.3 GHz H$92\alpha$ Line parameters} 
2000 Jun 09 & C & $-12.3$  &  226 & 7.1 & 1.1\\
2000 Jul 19 & D & $-12.3$ &  226 & 7.1 & 0.9 \\
2000 Sep 29 & D & $-12.3$  &   452 & 28.2 & 1.9\\
2000 Oct 03 & D$\rightarrow$A & $-12.3$ &  452 & 28.2 & 1.8\\
2001 Oct 09 & DnC & $-200.0$ & 452 & 14.0 & 1.8\\
2001 Dec 27 & D & $-200.0$ & 452 & 14.0 & 2.3\\
\cutinhead{1.4 GHz \HI\/ Line parameters} 
2002 May 13 & A$\rightarrow$B & $-40.0$ & 660 & 2.6 & 1.0\\ 
2002 Jun 06 & BnA & $-40.0$ & 660 & 2.6 & 1.2\\ 
\cutinhead{Archival 1.4 GHz Continuum parameters}
1984 Mar 11~$^b$ & CnB & N/A & N/A & N/A & 1.0 \\
1985 Dec 13~$^c$ & D & N/A & N/A & N/A & 2.4\\
\cutinhead{Archival 4.9 GHz Continuum parameters}
1984 Jul 19~$^b$ & DnC & N/A & N/A & N/A & 0.4 \\
\enddata
\tablenotetext{a} {Approximate time on source.}
\tablenotetext{b} {VLA project code AB254.}
\tablenotetext{c} {VLA project code AB333.}
\end{deluxetable}

\begin{deluxetable}{lcrc}
\small
\tablewidth{34pc}
\tablecaption{\HII\/ Region G357.63$-$0.06 Continuum Parameters\label{tab2}}
\tablecolumns{4}
\tablehead{
\colhead{Frequency}  & \colhead{Integrated Flux~$^a$} & \colhead{Beam} & \colhead{Size~$^b$}\\
\colhead{(GHz)} & \colhead{(mJy)} & \colhead{($\arcsec\times\arcsec $ [P.A.])} & \colhead{($\arcsec\times\arcsec$ [P.A.])}}     
\startdata
1.4~$^c$ & $69\pm 3$ & $13.6\times 11.4$ [$-76\arcdeg$]  & $8.0\times 5.8$ [$+111\arcdeg$] \\
1.4~$^d$ & $69\pm 2$ & $5.4\times 3.0$ [$-50\arcdeg$]  & $7.2\times 5.7$ [$+147\arcdeg$] \\
4.9~$^e$ & $82\pm 2$ & $12.8\times 8.7$ [$+47\arcdeg$]  & $7.2\times 5.9$ [$+119\arcdeg$] \\
8.3~$^f$ & $89\pm 3$ & $14.3\times 6.8$ [$+12\arcdeg$]  & $7.2\times 5.2$ [$+107\arcdeg$] \\
8.3~$^g$ & $88\pm 1$ & $3.1\times 2.1$ [$+15\arcdeg$] & $7.6\times 5.7$ [$+122\arcdeg$] \\
\enddata
\tablenotetext{a} {Integrated flux calculated after primary beam correction.}
\tablenotetext{b} {Deconvolved sizes from JMFIT.}
\tablenotetext{c} {Archival 1.4 GHz continuum data.}
\tablenotetext{d} {1.4 GHz continuum constructed from the line-free \HI\/ data.}
\tablenotetext{e} {Archival 4.9 GHz data.}
\tablenotetext{f} {Naturally weighted 8.3 GHz data with UV range up to 100 k$\lambda$.}
\tablenotetext{g} {Uniformly weighted 8.3 GHz data with UV range up to 350 k$\lambda$.}
\end{deluxetable}


\newpage

\begin{thebibliography}{}

\bibitem[Becker \& Helfand (1985)]{Becker1985} Becker, R. H., \& Helfand, D. J. \ 1985, Nature, 313, 115

\bibitem[Bitran et al. (1997)]{Bitran1997} Bitran, M, Alvarez, H., Bronfman, L., May, J., \& Thaddeus, P. 
\ 1997, A\&AS, 125, 99

\bibitem[Brogan et al. (2000)]{Brogan2000} Brogan, C. L., Frail, D. A., Goss, W. M., \& Troland, T. H. \ 2000, 
ApJ, 537, 875

\bibitem[Burton \& Liszt (1978)]{Burton1978} Burton, W. B. \& Liszt, H. S. \ 1978, ApJ, 225, 815

\bibitem[Burton et al. (2002)]{Burton2002} Burton, M. G., Lazendic, J. S., Yusef-Zadeh, F., \& Wardle, M. \ 2002, MNRAS,
submitted?

\bibitem[Caswell et al. (1980)]{Caswell1980} Caswell, J.  L., Haynes, R. F., Milne, D. K., \& Wellington, K. J. \ 1980, MNRAS, 
190, 881

\bibitem[Caswell et al.  (1989)]{Caswell1989} Caswell, J.  L., Kesteven, M.  J., Bedding, T.  R., \& Turtle, A.  
J.  \ 1989 Proc.  Astron.  Soc.  Aust., 8, 184

\bibitem[Cram et al. (1996)]{Cram1996} Cram, L. E., Claussen, M. J., Beasley, A. J., Gray, A. D., \& Goss, W. M. 
\ 1996, MNRAS, 280, 1110

\bibitem[Dame et al. (1987)]{Dame1987} Dame, T. M., Ungerechts, H., Cohen, R. S., de Geus, E. J., Grenier, I. A., 
May, J., Murphy, D. C., Nyman, L.-A., \& Thaddeus, P. \ 1987, ApJ, 322, 706

\bibitem[Dame, Hartmann, \& Thaddeus (2001)]{Dame2001} Dame, T. M., Hartmann, Dap \& Thaddeus, P. \ 2001, ApJ, 547, 792

\bibitem[Frail et al. (1996)]{Frail1996} Frail, D. A., Goss, W. M., Reynoso, E. M., Giacani, E. B.,  Green, 
A. J., \& Otrupcek, R. \ 1996, AJ, 111, 1651 

\bibitem[Helfand \& Becker (1985)]{Helfand1985} Helfand, d. J. \& Becker, R. H. \ 1985, Nature, 313, 118

\bibitem[Liszt \& Burton (1980)]{Liszt1980} Liszt, H. S. \& Burton, W. B. \ 1980, ApJ, 236, 779

\bibitem[Lockman, Pisano, \& Howard (1996)]{Lockman1996} Lockman, F. J., Pisano, D. J., \& Howard, G. J. \ 1996, ApJ, 472, 173

\bibitem[Milne (1970)]{Milne1970} Milne, D. K. \ 1970, AJP, 23, 425
 
\bibitem[Radhakrishnan (1972)]{Rad1972} Radhakrishnan, V., Goss, W. M., Murray, J. D., Brooks, J. W. \ 1972, 
ApJS, 24, 49

\bibitem[Shaver et al. (1985a)]{Shaver1985a} Shaver, P.  A., Salter, C.  J., Patnaik, A.  R., van Gorkom, J.  H. 
 \& Hunt, G.  C.  \ 1985a, Nature, 313, 113

\bibitem[Shaver et al.  (1985b)]{Shaver1985b} Shaver, P.  A., Pottasch, S.  R., Patnaik, A.  R., van Gorkom, J.  
H.  \& Hunt, G.  C.  \ 1985b, A\&A, 147, L23
 
\bibitem[Stewart et al. (1994)]{Stewart1994} Stewart, R. T., Haynes, R. F., Gray, A. D., \& Reich, W. \ 1994, ApJL, 432, L39

\bibitem[Shull, Fesen, \& Saken (1989)]{Shull1989} Shull, M. J., Fesen, R. A., \& Saken, J. M. \ 1989, ApJ, 346, 860

\bibitem[Weiler \& Panagia (1980)]{Weiler1980} Weiler, K. W., \& Panagia, N. \ 1980, A\&A, 90, 269

\end{thebibliography}
\end{document}